# AGILE, USER-CENTERED DESIGN AND QUALITY IN SOFTWARE PROCESSES FOR MOBILE APPLICATION DEVELOPMENT TEACHING

Manuel Ignacio Castillo López[1], Ana Libia Eslava Cervantes[2], Gustavo de la Cruz Martínez[2] and Jorge Luis Ortega Arjona[3]

[1]Posgrado en Ciencia e Ingeniería de la Computación, Universidad Nacional Autónoma de México, Mexico City, Mexico – ORCID: 0000-0002-2307-5860
[2]Instituto de Ciencias Aplicadas y Tecnología, Universidad Nacional Autónoma de México, Mexico City, Mexico
[3]Departamento de Matemáticas, Facultad de Ciencias, Universidad Nacional Autónoma de México, Mexico City, Mexico

*ABSTRACT*

*Agile methods in undergraduate courses have been explored in an effort to close the gap between industry and professional profiles. We have structured an Android application development course based on a tailored user-centered Agile process for development of educational digital tools. This process is based on Scrum and Extreme Programming in combination with User Experience (UX) approaches. The course is executed in two phases: the first half of the semester presents theory on Agile and mobile applications development, the latter half is managed as a workshop where students develop for an actual client. The introduction of UX and user-centered design exploiting the close relationship with stakeholders expected from Agile processes allows for different quality features development. Since 2019 two of the projects have been extended and one project has been developed with the described process and course alumni. Students and stakeholders have found value in the generated products and process.*

*KEYWORDS*

*Agile development, User-Centered Design, User Experience, Software development, Undergraduate teaching*

## 1. INTRODUCTION

In recent years, the *grupo Espacios y Sistemas Interactivos para la Educación* (ESIE) from the *Instituto de Ciencias Aplicadas y Tecnología* (ICAT) of the *Universidad Nacional Autónoma de México* (UNAM) has worked on improving their software development processes by including in them User Experiences (UX) design practices [1]. Also in recent years various authors have explored the inclusion of Agile processes in software development teaching [2 – 4].

The motives to include UCD in software development processes are related to Human-Factors software quality attributes, which have been identified as primary factors on user retention and success of a software project [5 – 7].

For this purpose, the grupo ESIE works on tools and spaces development for teaching or didactic activities, including gamification practices for *interactive experiences* building. The goal of products from grupo ESIE is to convey these experiences to users, in order to guide them through learning or reinforcement of knowledge. Specifically, the goal of the grupo ESIE by including UX design practices into its software process is to dispose of adequate tools for product design [1].

The way in which the group has approached the inclusion of Human-Factors in its software processes leans on the exercise of the process in courses its members taught as part of their academic activities at UNAM. In particular, the process of grupo ESIE has been applied repeatedly in the course *Programación de Dispositivos Móviles* (PDM) of the *Licenciatura en Ciencias de la Computación* (LCC) at the *Facultad de Ciencias* (FC), UNAM; a Bachelor in Science program on Computer Science. The application of the process introduces students to UCD and UX practices, while members of ESIE observe the effects of the process activities in the students' course projects.

This document presents a reflection of the results obtained so far by applying the grupo ESIE process in the PDM course, considering the means in which Human-Factors from an UX perspective change the priority of requirement analysis and design activities, and the impact the process has on students' projects.

The document is organized as follows: Section 2 presents similar approaches in which Agile methodologies have been applied into undergraduate courses; Section 3 provides a general description of the ESIE process; Section 4 describes the PDM course; Section 5 exposes the application of the ESIE process. Finally, Section 6 discusses project development beyond the PDM course and; Section 7 provides a conclusion, discussing the results obtained after several PDM courses.

## 2. AGILE PROCESSES INCLUSION IN UNDERGRADUATE TEACHING

Several proposals have emerged which explore the effects of incorporating Agile Processes in undergraduate courses. A recent work from Lara & Figueroa [2] applies *Extreme Programming* (XP) to a Software Engineering course, aiming to produce educational applications for mobile devices. This course is divided into phases, implied by XP in combination with teaching objectives: teaching requirements approach, pedagogical content design, user interface design, implementation, testing and delivery.

Their requirement acquisition strategy is based on User Histories, from which tasks are defined to conform the project's iterations. The solution design employs class diagrams, prototypes, and navigation maps. Tests consist of automated unit testing. From the release of this paper, the proposed course has offered the development of teaching applications to Argentine educational institutions, but it only has gotten to the requirements acquisition phase [2].

Tesei et. al [3] consider it necessary to incorporate Agile Processes into the teaching process to close the gap between professional profiles and industry needs. For this purpose, they design a security management software project to be developed in a course about software development.
The students are provided with development roles, while teachers act as Product owners. Some students are chosen as Scrum Masters, which do not directly perform coding activities. The course spans 16 weekly sessions, divided in 4 week-iterations. The first iteration focuses on work environment setup and solution design. Each session has a 4 hour length; the first two hours are dedicated to explore theory about Agile methodologies, while the next two hours are used to apply such theory. Scrum ceremonies are performed in each session [3].

As in the work by Lara & Figueroa [2], the design phase is based on User Histories. Students renegotiate with the Product owner the scope and deliverables for each iteration. At the time this paper was published, the course had not finished yet, but the students point out that the learning of Agile methodologies has been clear and useful through the project [3].

The work of Martín Gómez [4] is similar to that of Tesei et. al [3], as their goal is to apply Agile methodologies to encourage group work and digital competences, closing the gap between professional profiles and industry needs. The author points out that the educational system shares the primacy of people and interpersonal relationships with Agile methodologies, so in combination they should allow the development of autonomy, capabilities and abilities.

The proposal is to implement Scrum in combination with Kanban in a group of six students with the following objectives: develop autonomous learning competences, and develop social abilities such as communication, leadership and technological competences. Another similarity with Tesei et. al [3] is the distribution of Scrum roles: the professor acts as Product owner, while most students act as developers with a Scrum Master in the team [4].

## 3. GRUPO ESIE'S SOFTWARE PROCESS

The grupo ESIE's software process has been developed from the perspective of the UCD, leaning in UX design practices [1]. The process is divided in six phases that are iteratively carried out according to Agile strategies and principles. Its application in projects can be described by 3 stages which represent the main objectives of the phases each stage contains.
The six phases are described below, grouped in their correspondent stage [1]:

### 3.1. First Stage: Preproduction

This stage contains the *Definition* and *Design* phases. In this stage, the project is prepared: its objectives, scope and requirements are established, and finishes with the deployment of a high-fidelity prototype, a *minimum viable product*, which eventually evolves in the next stages until the final product is obtained.

*Phase 1) Definition.* Initial agreements are established with clients and stakeholders; their needs are documented, and the project objectives are defined.

*Phase 2) Design.* The product's central concept is defined, and the specification of experiences and characteristics of the product are described.

### 3.2. Second Stage: Production

Contains the *Development* phase, and shares the *Testing and validation* phase with the next stage. Tasks are selected to be developed in a short period of time by applying Agile methodologies. These tasks are performed on top of the results of the last iteration. Results can be assessed to obtain feedback for the next iteration, allowing timely identification of deficiencies on the design or its implementation, so they can be improved.

*Phase 3) Development.* The design obtained in the last stage is implemented on the results of the last iteration. Four general types of exceptions are recognized in this phase:

1. Limits on expertise from developers, who often are undergraduate students. Assistance and training is provided if such an exception occurs.
2. Limitations on tools. For example, a feature is limited by a selected library or runtime environment. Alternatives are explored, and the design is revised in case changes would be required by a tool alternative.
3. Limitations in the design. The design is revised in order to remove limitations without making major changes to the general product structure, just as when a tool requires a redesign.
4. The last exceptional task considers integration conflicts, which can happen when different tasks require the same artifact to be modified, so changes must be properly integrated.

All changes are reported and taken into consideration for the next iteration.
*Phase 4) Testing and validation.* The product is tested with target users employing usability testing techniques in a controlled environment.

### 3.3. Third Stage: Postproduction

This stage contains the C*losing* and *Feedback and maintenance* phases. The generated product is validated according to the specification agreed with the client. If the product's quality is determined to be within expectations, results are presented to the client who decides if the product is ready to be deployed or whether a new *Production* stage is required (this also depends on the scope and agreements).

*Phase 5) Closing.* The product is Deployed along with any other deliverables agreed with the client.

*Phase 6) Feedback and maintenance.* After the product is deployed, it is studied if the agreements with the client allows it. Modifications to enhance the generated experience can be considered for a next version of the project.

## 4. APPLICATION OF THE GRUPO ESIE'S PROCESS IN THE PDM COURSE

The activities of the grupo ESIE members include teaching. The PDM course is regularly taught by some of the group members. This course has a theoretical-practical approach, in which concepts around mobile applications development are presented during the first half of the semester, along with 3 or 4 example micro-projects. These concepts include some general mobile applications interface design and development with Android Studio. Students grading is mainly performed by assessing the software they develop, development performance, assignments, and a final project with an actual client.

Figure 1 shows students practicing Scrum though a low-fidelity prototype. Figure 2 shows students practicing Kanban for task management.

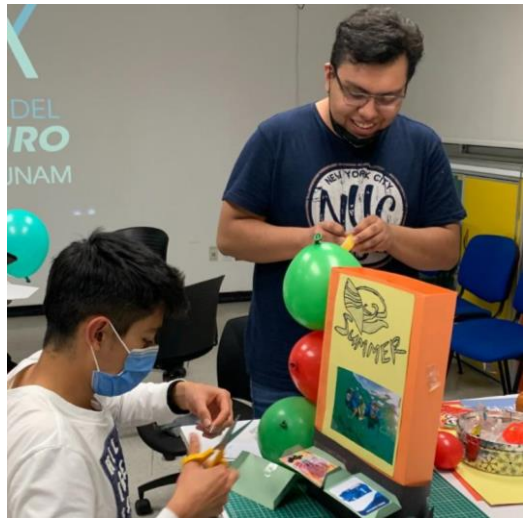

Figure 1. Students practice the Scrum methodology through a low-fidelity prototype.

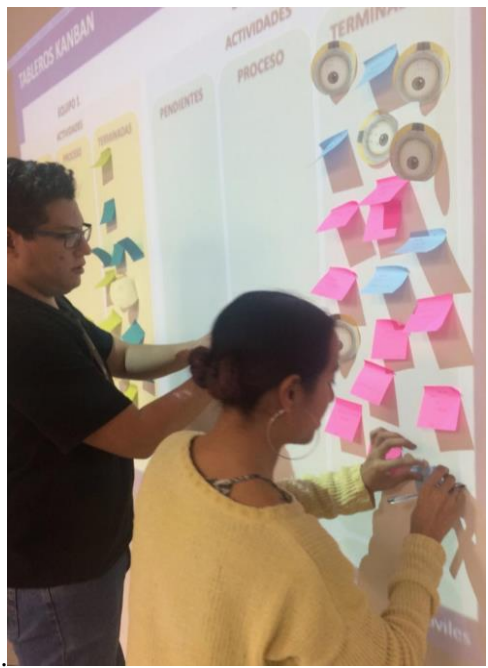

Figure 2. Students practice project task management through a Kanban Board

The second half of the course focuses on development of a software project with a client, who usually is a UNAM academic with a need around a mobile application. The project's objective is that students build a minimum viable product of the application specified by the client. The project applies the grupo ESIE's process, which has allowed multiple student groups to develop such applications, satisfying the client and users' highest priority requirements through UX and UCD activities in the process as a proof-of-concept product.

The students developed products are assessed with target users at the ICAT's *Aula del Futuro*, which is an specialized environment used for UX and usability testing with a focus on learning spaces and tools. Results from user testing are shared with the student's client, which are measured using *System Usability Scale* (SUS) [8]. SUS estimates the degree to which the product allows users to reach their goals under the expected use-context [5, 6, 9]. Figure 3 shows students performing a usability test at the *Aula del Futuro*.

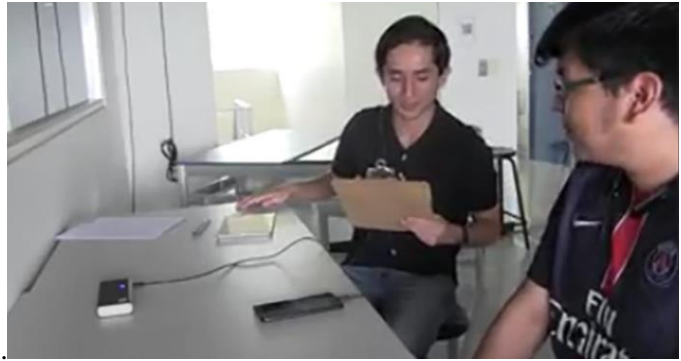

Figure 3. A student guides a user through a usability test

Project development is carried out by teams of 3 to 5 students, in which one of the team members has a *Facilitator* role, responsible for inspecting the task's progress though a Kanban board. Students are instructed to only change the tasks they are directly responsible for. The *Facilitator* also moderates the interaction between team members and assists them to get the necessary work resources. This role usually manages the team's software repositories.

The other team members choose at least one development role, such as Database Administrator or Designer; every team member including the *Facilitator* must perform programming tasks. LCC courses have (at least) two lecturers: a professor, who is administratively responsible for the course, and one or more assistants. In the PDM course, the professor has a Coach role within the students projects, providing assistance, orientation and feedback about their process implementation and results. Assistants have a Product owner role, reviewing the built product, orienting and providing feedback about the produced software artifacts.

### 4.1. Preproduction Stage

Following the 3 stages in which the six phases of the grupo ESIE's process develops, the process begins with the description of the students activities by the *Preproduction* stage in the PDM course, starting in the *Definition* phase, in which students must prepare an interview with the client for requirements acquisition, client needs, and expectations. Then the *Design* phase begins as the students prepare a first product proposal, to be presented to the client, and further refine their design.

The student's proposal includes a target user profile analysis for UCD tasks development through the *Persona* technique [10]. The user profile is obtained from questionnaires published online in forums and other websites, where target users are expected to be found. The proposal is built around the user's profile, client requirements and expectations. In particular students are requested to focus on navigation and user interface design. The proposal is prepared in a low-fidelity prototype, usually built on paper, which represents the most significant functionalities. From this first prototype, the proposal is further refined and a second prototype is produced. This second prototype is then presented to the client, along with the full proposal.

### 4.2. Production Stage

The *Development* phase starts with the observations made by the client about the student's proposal, for a next refining and task prioritizing round. The release cycle has a length between one and two months, with iterations of one or two weeks long.

The prioritization of tasks is reflected on a Kanban board, and an Android Studio project is prepared in a team controlled Git-based repository. Through *Continuous Integration* and *Continuous Delivery*, students carry out their programming tasks in individual Git-branches, and at the end of every iteration changes, these are integrated into a branch which represents the final build for the iteration.

### 4.3. Postproduction Stage

When the *Production* of the application is done, the generated product is *Tested and Validated*. The product should expose the viability of the proposal made by the students. To appraise the degree with which the product addresses the client's and user's needs, usability tests are performed, observing user interaction sessions in a usability laboratory. For these tests, students develop a series of assessment instruments and a test-session script, which dictates the activities to perform at the test. Particularly, those activities which the user must perform through the product, which should match with the most relevant functionalities identified. An exit questionnaire is also developed to collect the user's opinion on the effectiveness of the product, to allow them to perform the requested tasks. These questionnaires are structured as a User satisfaction survey. Observations of the tests, answers to assessment instruments and questionnaires are analyzed to produce recommendations, which should be addressed in a next version of the produced application.

The application of grupo ESIE's process in the PDM course finishes with the *Closing* phase. Test results are presented to the user along with the tested product. *Feedback and maintenance* can happen if new agreements are made between the client, grupo ESIE members, and the students, for a new version of the project, which could be part of the student's graduation activities.

## 5. GRUPO ESIE'S PROCESS ACTIVITIES IN STUDENT'S TASKS

Grupo ESIE's process defines some general activities to be performed along a software project, with specific inputs and required conditions for the described activities, as well their specific outputs. These process activities are derived from Scrum and Extreme Programming practices. Applying grupo ESIE's process in the PDM course's projects offers a familiar environment for the group members, who coach the students developing mobile applications and encompasses task planning and distribution, quality frameworks and UCD practices for the students to develop their applications on.

At the beginning of each project process phase, it is presented to the students the work strategies for the phase activities, their inputs, outputs and required conditions. The process activities are integrated into the PDM course as tasks and their outputs are graded. Students plan their tasks according to the current phase framework, and then, they practice the process activities.

It is important to note that grupo ESIE's process is actually a meta-process, while its tasks have required inputs and expected outputs. Many of the tasks, deliverables, number of iterations, and priorities are defined for each project, and each of their iterations. Actual tasks are basically instantiated on-the-fly according to the iteration goals; iterations can be seen as milestones which add-up the general project objectives.

By the course's end the students recognize various of the Agile principles in which grupo ESIE's process is based on, and the effects these principles have on their work. Students who continue developing applications have followed the learned practices, and built software artifacts with a UCD approach.

### 5.1. UX and DCU application in the PDM course

Commonly, UCD and UX activities are unknown to students at the beginning of the course. When the first tasks at the *Design* phase are performed, it is likely for the students to focus on functional requirements, while quality in use characteristics, user and client expectations are initially dismissed. As tasks develop and the identification of usability traits become more apparent to the students, they show a change of attitude towards UCD. The value of usability testing as a means for user quality assessment is usually highly regarded by students who finish the course.

### 5.2. Agile methodology application

Grupo ESIE's process is based on Agile methodologies with emphasis in collaboration principles with the client, users, and other stakeholders, who are seen as part of the development team. Agile methodologies offer a framework with conditions and mechanisms which enable collaboration and feedback required by user-centered approaches [5, 6, 11].

Along with UCD and UX activities, it has been frequently observed that task management, distribution and software repository management activities, such as versioning, become clearer by the end of the semester, even when students have had previous software product development experiences.

This may be the effect of the grading scheme, which not only grades the product, documentation, artifacts and development history in the project's software repository, but also includes task distribution and timeliness. Task management is graded through activity and history of the project's Kanban board at the end of each sprint. After reviewing each team's Kanban board, feedback is provided to the members of the team about their task management. As the project evolves, students refine their understanding of Agile concepts, challenges, and risks around team management.

The process is based on Agile methodologies, UCD and UX design. Its application on the PDM course preserves the basic concepts from all these software development approaches, which results in prioritization of the preproduction stage. At first glance, this appears to work against Agile principles such as *working software over comprehensive documentation* and *responding to change over following a plan* [12].

Several Agile projects show a literal interpretation of the Agile manifesto as a software methodology. Although the Agile manifesto is a high-level synthesis, which describes a framework, it lacks assessment tools and directives to properly define a methodology [11 – 13]. For example Tesei et. al [3] point out as an advantage of their proposal the reduction of documentation and the start of coding activities in early stages of the project.

Building software with minimal documentation usually has negative impacts on software lifecycle stages after its release. Agile methodologies proposed or promoted by the Agile manifesto authors have mechanisms for delaying or automatizing documentation production, to produce informal or non-conventional documentation, which is as useful but easier to manage than traditional documentation. Documentation is not de-prioritized nor removed from an Agile project [11, 14, 15].

PDM course's students who have used development strategies with minimal documentation management have found the preproduction stage as cumbersome. This is due to the fact that the first weeks of the course's project are dedicated to design activities, which usually involve documentation and low-fidelity prototype production. However, very few (if any) coding tasks. By the end of the course, most of these students understand the practicality of a more exhaustive design stage, since it offers better planning opportunities based on a well-structured design. Development begins from a prototype which can be evolved into the final product in about a month.

Furthermore, the produced design and documentation generated in the pre-production stages are essential for project extensions, after the PDM course's end of semester, when agreements are reached between the grupo ESIE, students and the clients.

Moreover the application of grupo ESIE's process in the PDM course requires autonomy from students. Their proposals and designs from the clients' requirements are developed into several work products, which are graded. This methodology, in which students try the theory to elaborate product proposals, is very similar to educational models being preferred, instead of classical models of concept and process memorization [4].

Grading of several work products (such as Kanban board changes, software repository changes through version control, design documents, prototypes, code quality and usability performance) allow punctual and rich feedback which students can use to reflect on their working strategies and process performance.

### 5.3. Application of usability testing

The implementation of usability testing with target users at a usability laboratory shows students how users react to both defects and features they like, and the impact such reaction has on their desire to keep using or to reject the product. Product assessment allows students to measure their solution's usability and show the client these results so conformance with his needs is further exhibited beyond the required functionality.

### 5.4. Other practical aspects in the PDM course

The importance of customers and stakeholders satisfaction has been highlighted in the grupo ESIE process by delivering software tailored to their needs and expectations. In order to achieve this in short release periods, specific aspects of the products are prioritized according to agreements and each project scope. Because these priorities vary with each project, such aspects are not explicitly described in the process.

These aspects are also presented to PDM course students. Mainly, the class of interactions and interfaces users expect on a mobile application, relevant considerations for mobile devices programming, privacy, and security aspects.

**Expected interfaces on mobile applications**. Although some mobile devices have very high pixel-density screens, usually these devices are physically small, and therefore, their screens are not suitable to display the same amount of information compared to larger devices, such as PCs or laptops. There are special considerations for tiny mobile devices, such as smartwatches, and other considerations for larger devices, such as tablet computers [16].

Users usually hold mobile devices in specific positions which should be considered when designing interfaces. For example: smartphones are usually held with a single hand, and users use their thumbs to interact with applications. Some users find it inconvenient to input large quantities of text by using in-screen keyboards [16].

**Considerations on Mobile environments**. Even the most powerful mobile devices are limited when compared to other computing platforms. Their data buses, operation frequencies, heat dissipation, power availability, and other aspects have limitations, which should be taken into consideration when designing software for this kind of computer devices. Otherwise slow, battery-hungry, or expensive cellular data applications may be delivered [17].

**Privacy and security**. The most popular mobile Operating Systems, Android and iOS, are POSIX-compatible systems, and inherit security features from the POSIX specification. Each application runs on a private sandbox, and while there are mechanisms to make two applications interact with each other, this must be done by an intentional design which must be clearly presented to users before/during installation or at startup [17].
Unfortunately it has become commonplace to abuse the closeness users share with their devices. Mobile devices accompany users through their daily activities. Many users use mobile applications to keep track of their health, physical condition, sleep cycles, their social life, tasks, appointments, and other activities.

This is commonly abused by applications and digital services providers, participating in the personalized advertisement market without users knowledge, consent or under unclear terms [18 – 20]. In the PDM course not only are ethical considerations about user data management presented, but also legal privacy frameworks are also briefly explored.

The two user data protection current laws in Mexico are discussed. These are the *Ley federal de protección de datos personales en posesión de los particulares* and the *Ley general de protección de datos personales en posesión de sujetos obligados.* The first regulates what kind of data, how data should be asked for, and when a private entity can ask it from other entities, while the second applies to public institutions which require the handling of user data in order to provide their services in a similar manner to the former law.

Other legal frameworks, such as Europe's *General Data Protection Regulation*, are briefly mentioned, so students are aware that different regions require software products to comply with different legal requirements.

Furthermore, techniques to write safe software are presented according to the course's project scope, such as SQL injection prevention, data encapsulation, defensive programming techniques and using ciphered channels over computer networks.

## 5.5. Software quality approach in the PDM course

The structure of the PDM course aims for the creation of high-fidelity prototypes, which should allow project's clients to assess whether a final version of the application allows them to achieve their goals, and if other requirements should be addressed, and the viability of achieving the goals in terms of time and resources.

As presented in previous sections, when the course project application is delivered, it is accompanied with the usability test results as a product quality indicator. Although not every quality dimension is assessed to the same extent, the course content and grading scheme encompass additional quality dimensions.

Software quality is defined as *the degree to which a software product meets its requirements, whose must precisely reflect stakeholders needs and expectations* [7, 11, 21]. In order to work with this definition, an aggregated quality taxonomy from standards and practices for Software Requirements Engineering is taken into consideration [7, 22, 23]:

- **Functional attributes**. These represent tasks and responsibilities expected from the product. The functions required by the client are prioritized by students under the supervision of the professor and assistants. The more relevant functions are chosen in accordance with the students' understanding of target users and client's expectations. Even when functions have limited operation options, they should be exploitable by users, who should get the expected results from each function.
  As a quality attribute, it is completely necessary that selected functions work in the final high-fidelity prototype. Otherwise, during the usability tests, the activities users perform may not align with expectations, which would be reflected on the users' satisfaction survey.

- **Interface attributes**. These are product characteristics which allow other software, systems, users or operators to interact with the product. In the PDM course, design is guided by well established mobile interfaces patterns. Touch-gestures and other appropriate input methods are considered to use the different functionalities that the product has to offer to their users. Interface quality is assessed in the usability tests.

- **Quality in use attributes**. These are qualities or restrictions that the product has to comply with. The selected functions have to operate within the client's defined margins or margins expected by users (using similar applications as a basis). Grading on design and product implementation considers closeness to stakeholders expectations.

- **Usability attributes**. Encompasses product characteristics which answer users' needs. This dimension is explicitly assessed by students and presented to the client on product delivery. The overall usability score and users' satisfaction survey are part of grading.

- **Human-Factor attributes**. These are characteristics that mitigate limitations on health and other conditions that may restrict the product's usability. The PDM course projects have not yet prioritized many Human-Factor characteristics. A future project may require a higher priority for these characteristics. They might be graded as part of the usability tests, verifying that usability is being promoted under the requested conditions. User satisfaction surveys can also indicate the degree to which the usability is being improved.

International Journal of Software Engineering & Applications (IJSEA), Vol.14, No.5, September 2023

## 6. PROJECTS CONTINUED BEYOND THE PDM COURSE

The PDM course has been taught yearly, using the described process since 2019. Two of the course projects have been continued after the semester, in which these projects ended. The students participate in the new version project as part of their graduation process.

The first project which had continuity is *Hocus Focus*, an application designed for people with Attention Deficit – Hyperactivity Disorder (ADHD). The application offers note taking functionality and mainly an habit agenda as shown in figure 4 (left). Habits are associated with daytime and an alarm (see right of figure 4), so ADHD users can keep a constant rhythm with their activities by following the alarms.

The usability test of *Hocus Focus* has an average usability score of 92.86%, as reported by users through the satisfaction survey. Observations from interactions and user's activity performance during test sessions pointed out some flaws in interactions, but did not report any major issues for users to achieve the requested session goals.

*Hocus Focus* has been awarded the *Premio a la Innovación UNAM 2019*, which is an annual prize given every year since 2018 to recognize innovative research and entrepreneurial projects and ideas. Hocus Focus was awarded the second prize in the social innovation category.

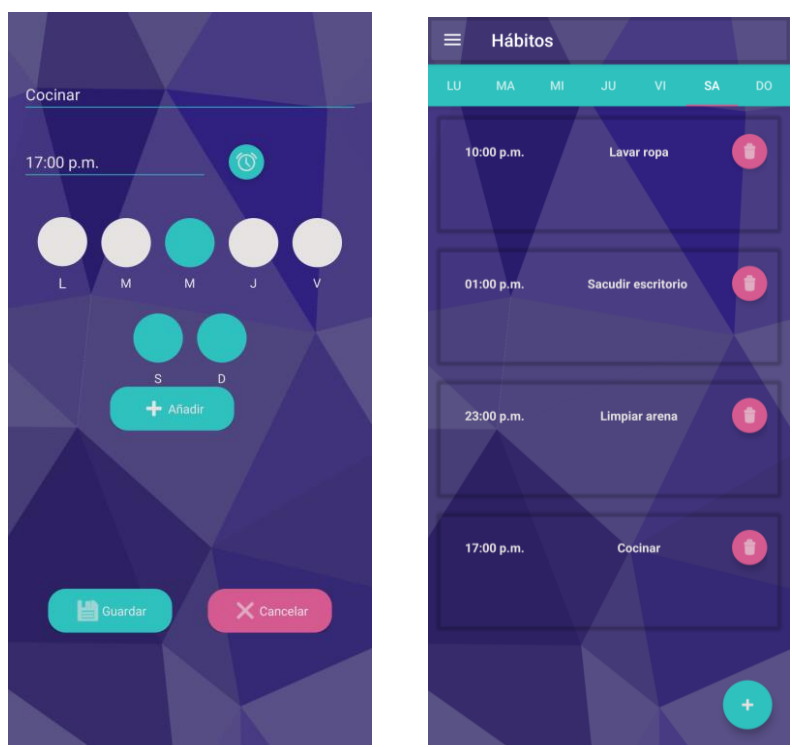

Figure 4. Some user interfaces of the *Hocus Focus* application. Left shows the *habit agenda* where user's habits are grouped by weekly occurrence. Right shows the habit registry where habits are setup for weekly occurrence at a specific time and a custom alarm ringtone.

Another project with continuity from the PDM course is *Polyx*, a polymers teaching-tool application for a course in a chemistry degree program taught at Facultad de Química, UNAM. The application contains theoretical concepts around polymers as shown in figure 5 (see left), and gamified self-reviews about the concepts (see right). Other academic dependencies at UNAM have expressed their interest in the application of the content/self-review structure of *Polyx* in different contexts.

According to the users' satisfaction survey, *Polyx* has a usability score of 89.6%. Performance of user activities show some issues about button labels and misuse of standard interface elements, such as scroll-bars and titles.

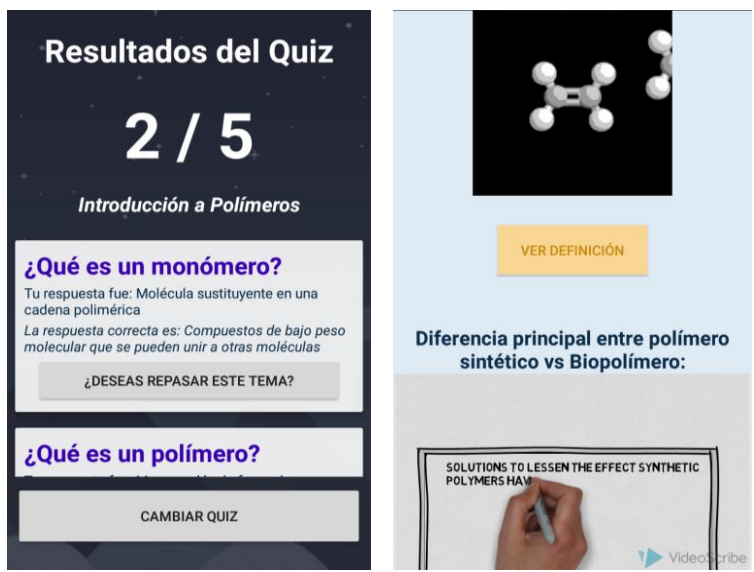

Figure 5. Some user interfaces of *Polyx*. Left shows the theoretical content organized topically. Right shows the result of a self-review test where it details which answers are right and which are wrong, providing the right answer and linking the theory to it.

The Dirección General de Bibliotecas (DGB) UNAM is interested on the interaction, content model, and self-review offered by *Polyx*, and reached an agreement with grupo ESIE to develop a similar application, named *DCide*, currently in development under the program UNAM-DGAPA-PAPIME number PE403222.

This application aims to help students to develop research competences through reliable sources identification among the enormous amount of information at their disposal through information technologies. To achieve this, *DCide* should help students to apply critical thinking during sources selection.

Although the *DCide* project has occurred outside a PDM course instance, from its conception it has been developed by course alumni using the medium-fidelity prototype from *Polyx*, as a reference for the DGB project. Students built a new proposal from the client's requirements and expectations, creating a new high-fidelity prototype, which has served as a base product to continue building *DCide*.

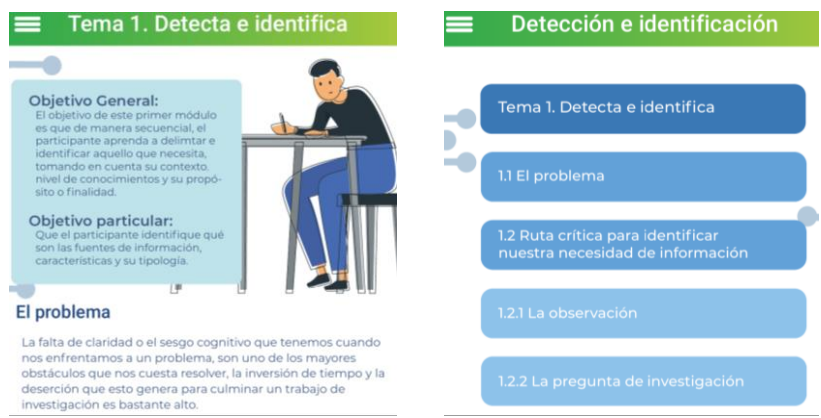

Figure 6. Some interfaces of the *DCide* application. Left shows the menu for one of the themes. Right shows the content structure for one theme.

The main differences between *Polyx* and *DCide* are:

- Thematic structure. *Polyx* distributes its content similarly to a polymers text book, while *DCide* distributes its content through five themes. Each theme arranges its content by chapters (see left of figure 6). *Dcide*'s five themes are: 1) Detect & identify, 2) Search & retrieve, 3) Select & asses, 4) Manage & analyze and 5) Use & tell.
- *DCide*'s content has more multimedia, tables and lists information structures.
- *Polyx* content is static. In order to change it, it is necessary to perform an application update. *DCide*'s content is more flexible and can be changed without necessarily performing a software update. There is a drawback where *DCide*'s content must be provided pre-formatted as an HTML document with all the necessary embedded CSS styles to show the content as desired.

## 7. CONCLUSIONS

Using the same principles as the related work presented in Section 2, the methodology applied in the PDM course combines practices from both Scrum and Extreme Programming. It has applied grupo ESIE's methodology annually since 2019, in collaboration with academics external to the group to develop minimal viable products for diverse domains.

Students who have taken the PDM course and later incorporated into the group's projects have shown an ease to structure and manage their activities, and a reflective attitude towards the project goals. Moreover, the development experience from the course helps them grasp a clearer vision of the activities they perform within the project, including their requirements, expected results and deliverables. Altogether, this attitude enables a collaborative, fluent, and reflective evolution of the projects.

This also has provided feedback to further refine the process, identifying ambiguous activities on it. Furthermore, the application of the process has helped students to practice UCD and to build Android applications, which satisfy their requirements, that should match the most relevant characteristics expected by the client, and which are presented to users through usable interfaces. Students also collect feedback from users to identify defects and opportunities that should be addressed in future versions.

Students also have shown the capacity to recreate UCD tasks in other projects and develop them by iterative and incremental means. On projects with continuity, the resulting project has become more complete and more correct in respect to the client's goals. Quality in use aspects are carefully managed, any modification to the user interaction has to be designed, implemented, and tested as specified by the grupo ESIE's process.

**ACKNOWLEDGEMENTS**

The authors would like to thank the academics and students who have made an effort for the development of projects derived from the PDM course. In particular, we would like to thank Lupita Vásquez Fabela for being the client for *Hocus Focus*, an application developed by the students Adrián F. Vélez Rivera, Aide I. García Hernández, Jesús M. Colín Torres, and Raúl Ascencio Bolio.

Also we would like to thank Dr. Yara C. Almanza Arjona for being the client for *Polyx*, developed by the students Joshua J. Pedrero Gómez, Mauricio Araujo Chávez, Paola Vázquez Rizo, and Uriel Rosas Franco. *Polyx* has continued its development by the student Jesús M. Colín Torres.

*DCide* is being developed with the support of the UNAM-DGAPA-PAPIME project number PE403222 "*El aula del Futuro en la Biblioteca Central*", developed by the students Lilián Ramírez Mekler, Derek Almaza Infante, and Vianey A. Borras Pablo.

Special thanks to the organizers of the *International Conference on Computer Science and Software Engineering* (CSSE 2023) for the invitation to present an improved version of our work exposed at CSSE 2023 [24].

International Journal of Software Engineering & Applications (IJSEA), Vol.14, No.5, September 2023

## AUTHORS

**Manuel Ignacio Castillo López**, Universidad Nacional Autónoma de México, Master in Computer Science. Sc.D student in the field of Software Process and Product Quality and Professor Assistant (PA) at FC, UNAM. Ignacio has collaborated with grupo ESIE since 2016 in product development and process evolution. He also has been PA for the PDM course on several occasions since 2019.

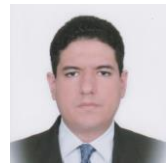

His interest in software quality comes from more than 10 years as a professional software developer and project manager, having encountered many of the challenges faced in the software industry and their hard outcomes, including layoffs and business cease of operations.

**Ana Libia Eslava Cervantes**, Universidad Oberta de Catalunya, Master in Multimedia Desing. Member of the grupo ESIE, ICAT, UNAM since 2015 and co-founder of the Aula del Futuro project,which aims to create a replicable model of a computer assisted teaching space. Libia is also the PDM course's professor.

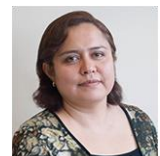

 The Aula del Futuro model not only offers a computar-based framework but also teaching practices to make an efficient use of digital tools. The model already has several implementations around Mexico and Latin America. Her work focus is on pedagogical innovation; design, development and assessment of interactive teaching environments with an UCD approach and of teaching strategies.

**Gustavo de la Cruz Martínez**, Universidad Nacional Autónoma de México, Sc.D in Computer Science. Founding member of the project of the Aula del Futuro. Professor at UNAM's FC since 2002 and full-time Academic at ICAT, UNAM since 2005. His areas of interest are human-computer interaction, user modeling and user experience. 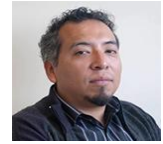

His current lines of research are: Methodologies for the design and evaluation of user experience, Cognitive user modeling, Interactive spaces and non-formal education.

**Jorge Luis Ortega Arjona**, Full-time Tenured Professor in the Department of Mathematics, Faculty of Sciences, Universidad Nacional Autónoma de México (UNAM). B.Eng. degree in Electronic Engineering, an M.Sc. degree in Computer Science from UNAM, and a Ph.D. degree in Computer Science from the University College London, in 2007. 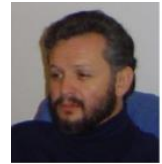

Author of five books, and more than 70 articles. His current research interests include parallel software design, parallel processing, object-oriented programming software patterns, and software design and architecture.